\pdfoutput=1
\documentclass[11pt,twoside]{article}  
\usepackage{graphicx}
\usepackage{asp2006}
\usepackage{adassconf}

\begin{document}   

%
%

\paperID{O5.3}

%

\title{LSST and the Dark Sector: Image Processing Challenges}

%
%
%
%
%

\markboth{Tyson, et al.}{LSST Multi-Fit Image Processing}

%
%
%
%

\author{J.~A. Tyson\altaffilmark{1}, C. Roat\altaffilmark{2}, J. Bosch\altaffilmark{1}, D. Wittman\altaffilmark{1}}
\altaffiltext{1}{Physics Department, University of California, Davis, CA, USA}
\altaffiltext{2}{Google, Mountain View, CA, USA}

%

\contact{Tony Tyson}
\email{tyson@physics.ucdavis.edu}

%
%
%

\paindex{Tyson, J.~A.}
\aindex{Roat, C.}
\aindex{Bosch, J.}
\aindex{Wittman, D.}     

%



\begin{abstract}          
Next generation probes of dark matter and dark energy require high
precision reconstruction of faint galaxy shapes from hundreds of
dithered exposures. Current practice is to stack the images. While
valuable for many applications, this stack is a highly compressed
version of the data.  Future weak lensing studies will require
analysis of the full dataset using the stack and its associated
catalog only as a starting point. We describe a ``Multi-Fit" algorithm
which simultaneously fits individual galaxy exposures to a common
profile model convolved with each exposure's point spread function at
that position in the image. This technique leads to an enhancement of
the number of useable small galaxies at high redshift and, more
significantly, a decrease in systematic shear error.
\end{abstract}

%
%

\section{Probes of dark energy and dark matter}
Dark energy affects the cosmic history of the Hubble expansion H(z) as
well as the cosmic history of mass clustering. If combined, different
types of probes of the expansion history and structure history can
lead to percent level precision in dark energy parameters. This is
because each probe depends on the other cosmological parameters or
errors in different ways. These probes range from cosmic shear, baryon
acoustic oscillations, supernovae, and cluster counting -- all as a
function of redshift z. Using the CMB as normalization, the
combination of these probes will yield the needed precision to
distinguish between models of dark energy (Zhan 2006).

Next generation surveys will measure positions, colors, and shapes of
distant galaxies over such a large volume that the resulting
stochastic (random) errors will be very small. It is necessary to
control and reduce the systematic errors to even lower levels.  There
are two primary systematic errors which can influence the data:
Photometric Redshift errors, and Weak Lens Shear errors.  The work to
date has employed highly idealized data models. Here we describe some
of the image processing challenges associated with reconstruction of
the galaxy images from many dithered exposures.

With its capability to go deep, wide, and fast, the LSST will yield
continuous overlapping images of 20,000 - 25,000 square degrees of
sky.  The baseline exposure time is 15 seconds, and each ``visit" to
a single patch of sky will consist of two such exposures separated
by a 2 sec readout with the shutter closed. In order to meet the
science goals, six bandpasses ($u, g, r, i, z,$ and $y$) covering
the wavelength range 320-1050 nm will be used.  The system is
designed and will be engineered to yield exquisite astrometric and
photometric accuracy and superb image quality. The telescope and
camera optics and the detector combine to deliver 80\% energy
within a 0.2 arcsecond pixel over the full 10 square degree field
and full wavelength range. This LSST survey will take ten years to
complete. In a ten-year survey, the LSST will make more than five
million exposures. In current simulations, the sky is tiled with a
fixed spherical covering of circular fields. This overlap leads to
a significant fraction of area which is observed twice as
frequently as the average. In practice, the position of each visit
will be varied continuously across the sky to average out this
extra exposure.

How is the precision of shear measurements of distant galaxies in weak
lensing tomography affected by ground-based seeing? Galaxy shape
measurement depends on three parameters: galaxy size, delivered PSF,
and limiting surface brightness. New ground-based telescopes are
routinely delivering 0.4-0.7 arcsec FWHM imaging without adaptive
optics. Clearly there are unique advantages in space for UV or IR
imaging.  Galaxies at 25 mag have mean half-light radius ~ 0.4 arcsec
and FWHM $\sim 0.8$ arcsec.  Angular sizes of galaxies change with
redshift due to a number of effects including the cosmological
angle-redshift relation, luminosity evolution, and surface brightness
dimming. The net effect is a plateau over a range of z, out to z=3
(Cameron and Driver 2007). At the low surface brightness reached in
hundreds of LSST exposures, typical galaxies at redshift $z<3$ can be
resolved sufficiently to measure their ellipticity. This is shown in
Figure 1.  One must convolve with the PSF and ask if the ellipticity
can be measured. Galaxies have a large intrinsic ellipticity (rms
$\sim 0.3$), and it is most important to have many of them in order to
average down the shot noise of this intrinsic ellipticity.  At 28-29
mag per sq. arcsec ground based seeing is sufficient to measure the
large ellipticities of 40-50 galaxies per square arcminute to the
required $z<3$ redshift limit for tomography. However, it is crucial
that shape systematics are minimized.

\begin{figure}
  \center
  \includegraphics[width=0.7\textwidth]{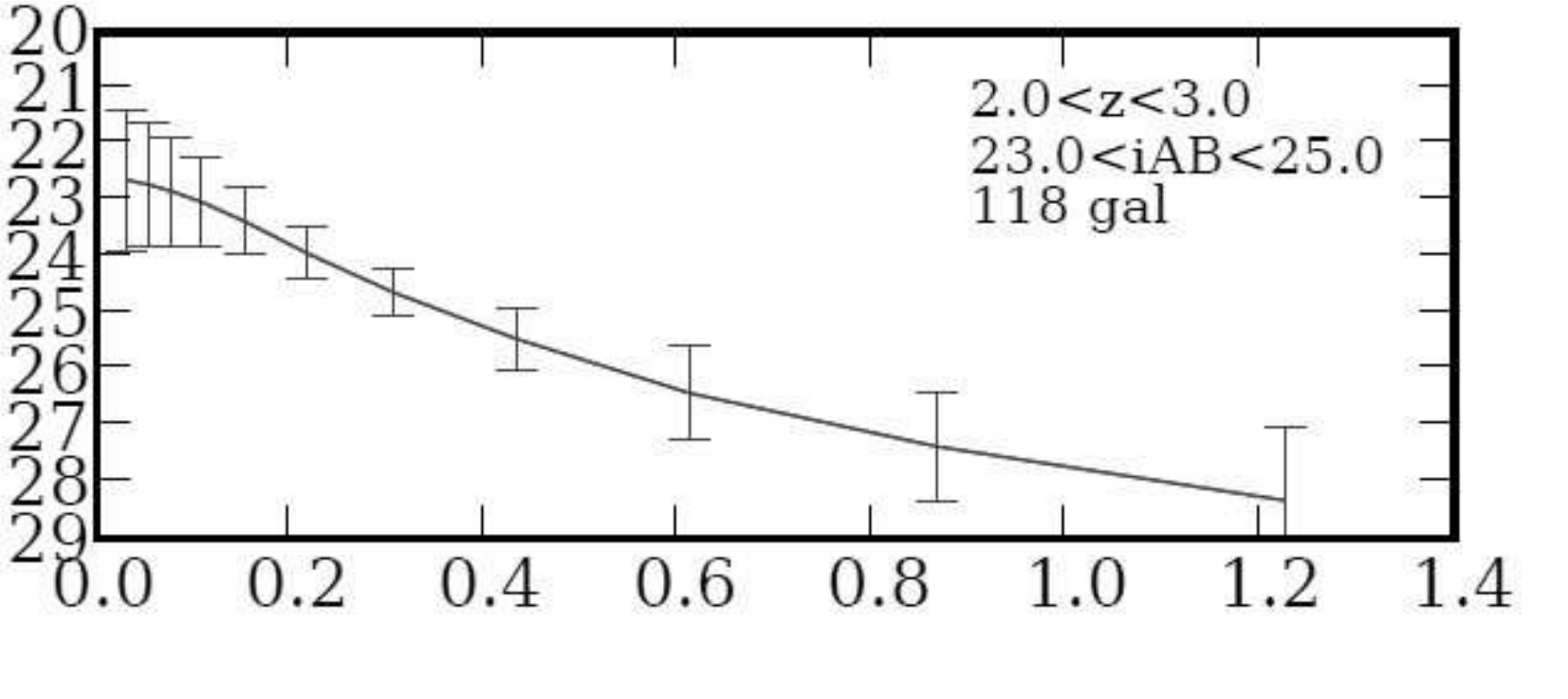}
  \caption{Galaxy surface brightness vs radius (arcsec) in one redshift bin
    from z = 2 - 3 for $23 < i < 25$ AB mag. This plot is from HST
    ACS imaging and ground based spectroscopy.  At the 28 $i$ and 29 $r$
    mag/sq.arcsec limit of the LSST survey most galaxies at $z<3$ are
    sufficiently resolved in 0.6 arcsec FWHM seeing to reconstruct their
    ellipticity. (courtesy H. Ferguson)}
\label{fig:O5.3_1}
\end{figure}

\section{Weak lens shear measurement}
Background galaxies are mapped to new positions on the sky by
intervening mass concentrations, shearing their images
tangentially. First detected in 2000 (Wittman, et al. 2000), the full
3-D cosmic mass distribution creates statistical correlations in
galaxy shapes called ``cosmic shear."  Systematic errors in either
redshifts or shear affect the precision obtainable for dark energy
parameters and are classified as either multiplicative or additive.
There is some level of self-calibration, especially for multiplicative
errors, i.e. the level of error can be obtained from the data and
marginalized over without severely compromising the cosmological
constraints. Additive errors do not have this property.

Multiplicative errors are also known as shear calibration errors, and
arise from the convolution of a galaxy's true shape with the isotropic
part of the point-spread PSF, which dilutes the shear by some factor
which depends on the relative angular sizes of the galaxy and the PSF.
Therefore multiplicative errors will be a function of redshift (more
distant galaxies appear smaller) and of position on the sky (the
ground-based PSF depends on the atmosphere).  Additive errors, or
spurious shear, arise from the anisotropic part of the PSF and are
position-dependent but not redshift dependent, except perhaps
indirectly, if the PSF is a function of source color.

\section{Shift-and-stare imaging}

If a large number of exposures are taken of a field on the sky it is
possible in principle to separate spatial defects on the imager from
the true scene on the sky.  Shift-and-stare imaging was developed in
the early days of CCD imagers for this purpose (Tyson 1986).  There
are a variety of algorithms for recombining the sub-images in this
data cube into a master co-added image.  The original technique used
median averaging a pixel of fixed sky location up the registered stack
of sub-images, but care must be taken not to introduce
correlations. Using sinc interpolation rather than simple weighted
neighbor pixel interpolation one can decorrelate noise on adjacent
pixels in the co-added image, making it possible to estimate
statistical significance.  Shift-and-stare is the method of choice
currently in all wide field deep imaging. However, it probably has
outlived its usefulness. While it is convenient from a storage and
computation point of view to compress the data cube to a single
co-added image, important information is lost particularly if image
quality or effective exposure varies between sub-images.

\begin{figure}[t]
  \center
  \includegraphics[width=0.7\textwidth]{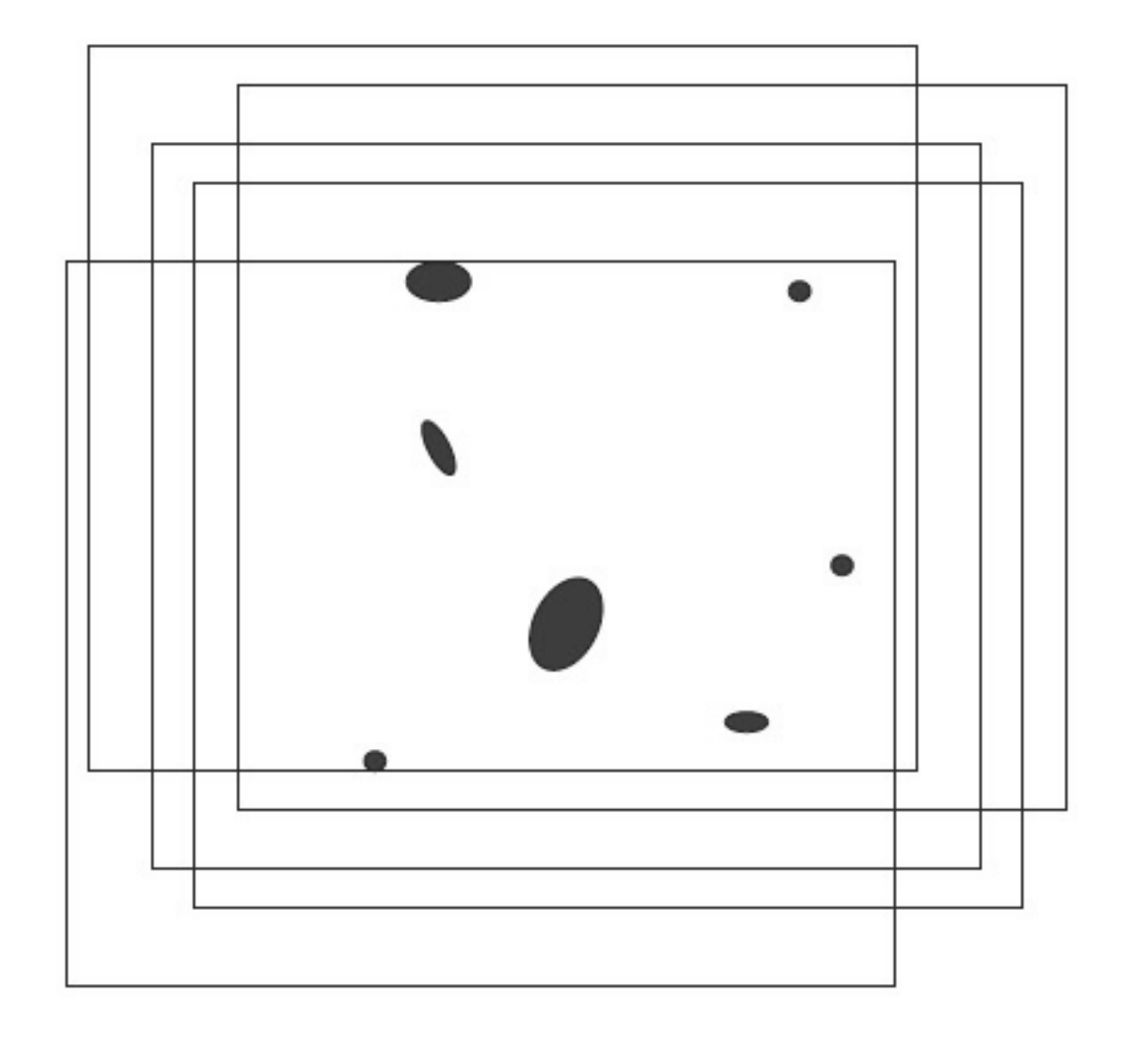}
\caption{Shift-and-stare: Multiple exposures disregistered on the sky
  contain information about the objects as well as information about
  defects on the imager.  Stars and galaxies are disregistered between
  exposures, but systematic errors in the CCD are registered in each
  frame.  Processing with a 'superflat' can remove most of the CCD
  based defects, and then registering the stars an co-adding generates
  a deep defect-free image.  Subtle problems can occur if the PSF is
  different in each image.}
\label{fig:O5.3_2}
\end{figure}

\section{Reconstructing galaxy images: co-addition vs Muti-Fit}

Several algorithms have been suggested to beat down PSF systematics
using multiple exposures of the same field.  The naive use of such
data would be to construct a single stacked image with higher
signal-to-noise, and then measure the shear correlation function by
averaging over all pairs of galaxies.  This requires pixel
interpolation, which can lead to systematics and correlated noise.
Generally the stack algorithms combine sub-images with different PSF,
and information is lost.  Moreover, there is generally a discontinuous
PSF change in the stack image at locations of edges of the
sub-images. This creates PSF variations on the stack image that are
hard to model.  As a result the stack method does not provide the
desired accuracy for image analysis algorithms which are sensitive to
spatial variations in PSF.

We propose analyzing the full ``data cube" by fitting, for each galaxy,
a single model which, when convolved with the $N$ different PSFs, best
matches the $N$ measurements of that galaxy (the MultiFit method).  This
means that PSF misestimation, which is strongly decorrelated from
image to image, behaves more like a random error for that galaxy,
rather than a systematic error. LSST will have hundreds of dithered
images per filter band per sky patch, and there will be about 2000
overlapping (dithered) 10 square degree sky patches per bandpass.  It
is desirable to use all the information in those overlapping data
cubes.

The best current methods reach a shear calibration accurate to 1\%. In
principle LSST can do 20 times better because LSST will have hundreds
of exposures, each with an independent shear calibration. Current
shear analysis operates on the co-added deep image. A new method,
Multi-Fit, does a superior job of estimating the true shear of a
galaxy by fitting a model to all images of it in the stack of $N$
exposures.

\subsection{Multi-Fit}
 
We describe a method for fitting the shapes of galaxies that have been
imaged in multiple exposures. Instead of the traditional approach of
co-adding many exposures to produce a single image for measurement,
this method simultaneously analyzes all individual exposures to
determine the galaxy shape and size that best fits all images of a
single galaxy in a noise-weighted fasion. This process effectively
uses knowledge about the PSF of individual exposures, taking advantage
of the detailed information present in highly resolved images, while
still extracting the limited information available in images with
poorer resolution. A PSF map is made for each image, by fitting all
the stars. The simultaneous fit is performed using a maximum
likelihood technique that combines the likelihoods calculated from
each individual exposure. First, a parameterized model for a galaxy
radial light profile is chosen. The model is convolved with each of
the PSF models measured from the individual exposures. The final,
convolved light distributions are compared to the data pixels for the
galaxy images on each individual exposure to determine a
likelihood. The fitting procedure adjusts the parameters of the input
model until the likelihood is maximized, resulting in a best-fit model
of actual galaxy shape prior to the effects of PSF smearing.

There are several advantages to using a procedure that fits multiple
exposures. First, errors that are made in PSF estimation in each
exposure are treated as random errors, and these errors are propagated
into the statistical error calculated during the fitting
process. Thus, these errors are determined directly for each
individual galaxy, rather than being an unknown systematic error.
Compared to interpolating PSF estimation on a co-added image, this
also reduces any spatial correlation introduced by PSF mis-estimation
in a given region of sky. A second advantage of this method is that
the PSF interpolation is done on each separate exposure, where the PSF
is expected to vary smoothly. Other methods interpolate on a co-added
image, which has been made using many exposures that have been
dithered relative to each other. The spatial variation of the PSF on a
co-added image is not smooth near the boundaries of the underlying
chips, making accurate interpolation more difficult.

Another advantage, specific to any technique that uses fitting, is
that prior information can be directly incorporated into the fit. The
choice of an underlying galaxy shape profile is one such piece of
information. Parameters of the galaxy-model or the PSF-model can be
constrained with additional terms in likelihood calculation. For
example, if the PSF determination is uncertain, those uncertainties
can be used in the fit and directly propagated into the final
measurement error. Priors based on the high S/N features of an object
in the stacked deep image are useful. The centroid of objects is taken
from the stacked image in our tests shown below and is not allowed to
vary from sub-image to sub-image in the data cube.

The following plots illustrate how well the ellipticity of galaxies of
different magnitudes and sizes can be measured. Below a pre-seeing
size of 0.5 pixels, fitting becomes unstable due to the small
size. Above a FWHM of 10 pixels, a minimum error is reached for a
fixed magnitude. A joint fit to the size and magnitude dependence of
the error, between 0.5 and 10 pixels, gives the expected statistical
dependence based on signal-to-noise, thus demonstrating the extreme
robustness of this technique.

\begin{figure}
  \center
  \includegraphics[width=0.45\textwidth]{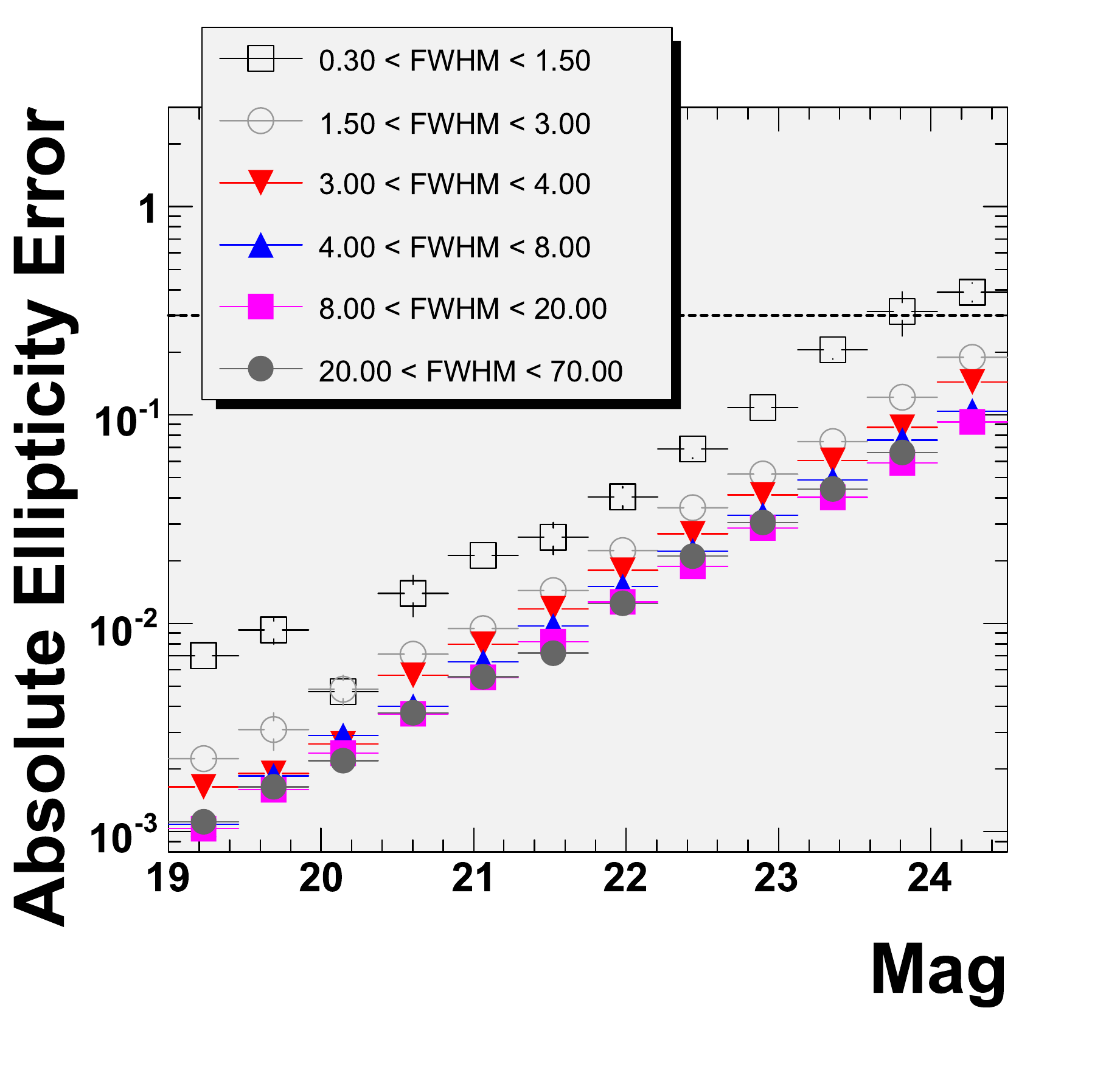}
  \includegraphics[width=0.45\textwidth]{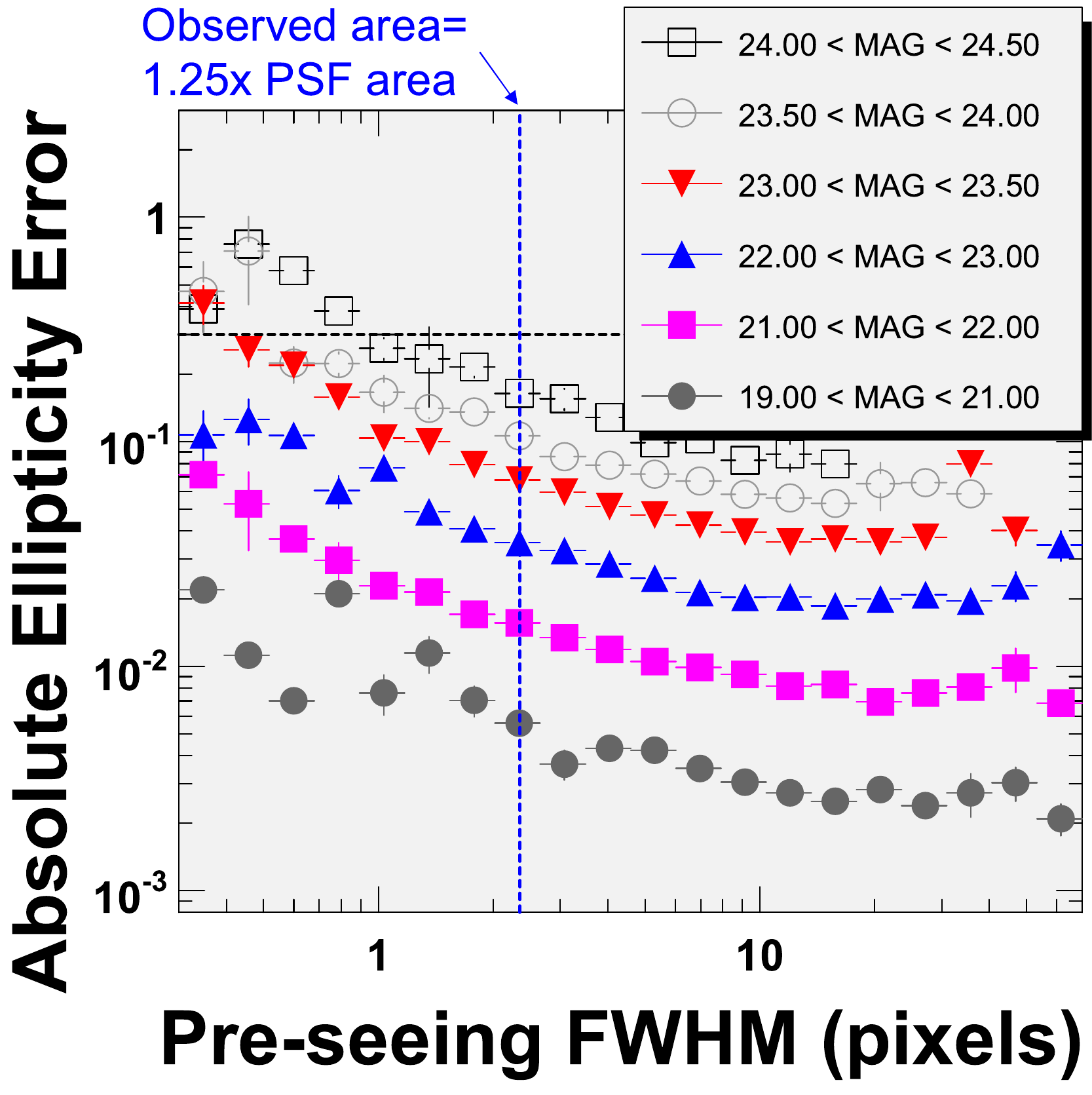}
\caption{Dependence of galaxy shape measurement error on magnitude
  (left) and galaxy size (right) in a simulated exposure cube. The
  dotted horizontal line indicates the level of shape noise - the
  intrinsic distribution of galaxy shapes. The magnitude variation is
  due to the higher signal-to-noise measurement possible with brighter
  objects. The variation of error with size shows that larger objects
  are measured better up to a magnitude-dependent noise floor. The
  vertical line is the level below which many current methods become
  unstable - when the observed area is 1.25x the PSF area.}
\label{fig:O5.3_3}
\end{figure}

The statistical figure of merit for a weak-lensing survey is the
effective number of galaxies for which shapes have been measured.  By
measuring ever smaller and fainter galaxies, a survey can dramatically
increase galaxy sample size, but at the cost of using noisier
measurements. There is a trade-off between increased shot noise plus
lower systematic shear error at the bright end and decreased shot
noise (due to the large number of galaxies) and susceptibility of PSF
sysytematics at the faint end. Many current methods for shape
measurement become unusable when observed objects have sizes close to
the PSF size. Often, galaxies observed to be less than $\sim 1.25$
times the area of the PSF are discarded. With fitting techniques, this
limit can be reduced and galaxies can be measured almost down to the
size of the PSF. The variance of a shape measurement decreases as the
square-root of the pre-seeing area for small galaxies. Since the
number of galaxies increases with the decreasing angular size, the
rapid increase in sample size can compensate for increased
noise. Consequently, the effective number of galaxies of a survey can
be substantially increased by recovering barely resolved galaxies. The
following figure depicts the relative increase in a survey's effective
sample size as galaxies less than 1.25 times the PSF area are
included.

This algorithm uses all information in the images, weights
better-seeing images appropriately, and handles image boundaries.  PSF
on a stacked image changes abruptly at a sub-image boundary. Each
sub-image PSF has less structure than the stacked image PSF, and this
approach thus transforms some systematics into random errors.

\begin{figure}
  \center
  \includegraphics[width=0.5\textwidth]{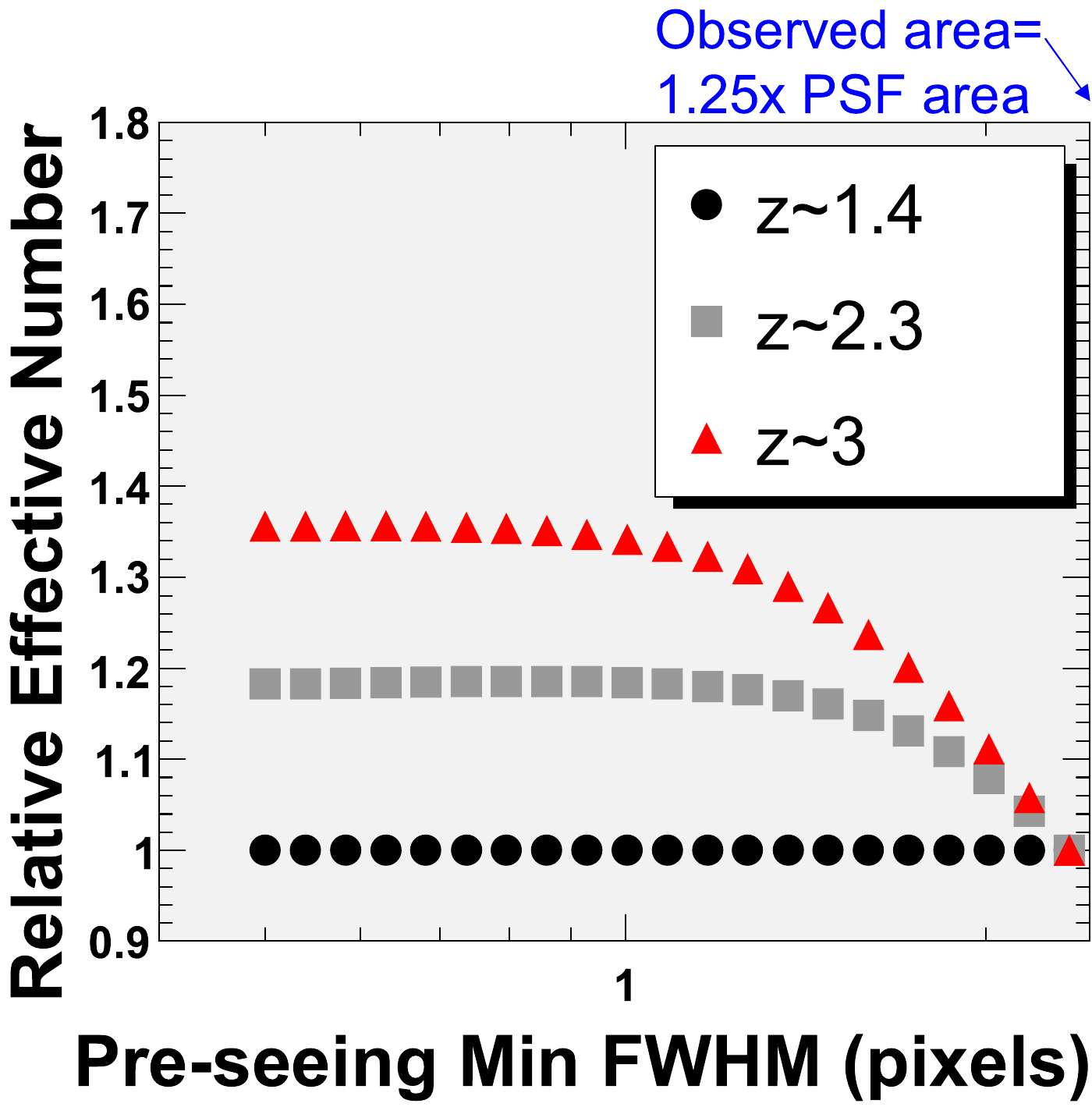}
\caption{Effective galaxy sample size relative to a sample with a
  cutoff of observed galaxy size 1.25 times the PSF size (2.35 on the
  x-axis, in these units). At low redshift, there are few galaxies
  present to add to the sample, while at higher redshift, there is
  some gain in measuring galaxies down to a cutoff of $\sim 1$.}
\label{fig:O5.3_4}
\end{figure}

\subsection{Computational challenges and R\&D to be done}

Currently, galaxies and PSFs are modeled sums of Gaussians, so
convolutions are fast. Real galaxies are not Gaussian, and an upgrade
to more realistic models has begun.  The current algorithm
requires 1 sec per galaxy for data cube of 20 images, with no speed
optimization yet, on a 2 GHz desktop. For the 5 million images LSST
will obtain, a rough extrapolation of the existing Multi-Fit runs
suggests over $10^{22}$ floating point operations.  This is
competitive with the computational requirements for the LSST image
differencing transient pipeline. The new code is being written in C++
and Python. It will be necessary to quantify the improvement of
Multi-Fit over stacking for various science cases (weak 
lens shear, photometry).  It will be particularly useful to extend
fitting to include other quantities: magnitudes, colors, etc., or to
use them as priors for single-band galaxy reconstruction. Finally, we
will pursue speed optimization and extensive Monte Carlo tests. We
propose to use Multi-Fit in full shift-and-stare Monte Carlo
simulations of LSST sky tiling operations including PSF systematics.

\end{document}